\documentclass[aps, pre, onecolumn, amsmath, superscriptaddress,showkeys,showpacs]{revtex4-2}
\usepackage[utf8]{inputenc}
\usepackage[title]{appendix}
\usepackage{amsmath}
\usepackage{graphicx}
\usepackage{amssymb,color,tikz,multirow}
\usepackage{mathrsfs}
\usepackage{fontenc}
\usepackage{float}
\usepackage{amsthm}
\usepackage{hyperref}[colorlinks=true, linkcolor=blue, citecolor=blue]
\usepackage{algpseudocode}
\usepackage{algorithm}

\usepackage{amsfonts}
\usepackage{booktabs}

\graphicspath{ {./} }

\begin{document}

\title{A First Principles Approach to Trust-Based Recommendation Systems in Social Networks}
\date{Feb, 2025}
\author{Paras Stefanopoulos}
\email{paras.stefanopoulos@outlook.com}
\affiliation{School of Computing, The Australia National University, Canberra, Australia}
\author{Sourin Chatterjee}
\email{sourin.chatterjee@univ-amu.fr}
\thanks{Corresponding author}
\affiliation{Department of Mathematics and Statistics, Indian Institute of Science Education and Research, Kolkata, West Bengal 741246, India}
\affiliation{Institut de Neurosciences des Systèmes (INS), UMR1106, Aix-Marseille Université, Marseilles, France}
\author{Ahad N. Zehmakan}
\email{ahadn.zehmakan@anu.edu.au}
\affiliation{School of Computing, The Australia National University, Canberra, Australia}


\begin{abstract}
This paper explores recommender systems in social networks which leverage information such as item rating, intra-item similarities, and trust graph. We demonstrate that item-rating information is more influential than other information types in a collaborative filtering approach. The trust graph-based approaches were found to be more robust to network adversarial attacks due to hard-to-manipulate trust structures. Intra-item information, although sub-optimal in isolation, enhances the consistency of predictions and lower-end performance when fused with other information forms. Additionally, the Weighted Average framework is introduced, enabling the construction of recommendation systems around any user-to-user similarity metric. All the codes are publicly available on GitHub \cite{Paras2023}.
\end{abstract}

\maketitle

\vspace{-2em} 
\section{Introduction}
Recommendation systems have become an essential component of online platforms, helping users navigate vast amounts of information and make decisions based on personalized suggestions. In social networks, these systems play a key role in suggesting products, content, friends, or pages to users based on their preferences, behaviors, and social connections. Social networks provide a rich source of information on user interactions and preferences, which can be leveraged to create more accurate and personalized recommendations. This type of recommendation system considers not only the past behaviors and preferences of users but also the behaviors and preferences of their connections or friends in the network \cite{campana2017recommender}.

Trust is a crucial concept in social networks and plays a vital role in recommendation systems~\cite{donovan_smyth_2005, andersen2008trust, papakyriakopoulos2020political, zehmakan2023rumors, htun2013trust,liu2023fast, fernandes2020isabela}. The abstraction of social connections as 'trust' is essential because it allows the system to consider implicit behaviors and preferences of users and their connections. For example, a user liking or sharing content, or having many mutual connections, can indicate a level of trust or shared interest, even if the users are not explicitly connected or have not directly interacted. This abstraction helps in capturing the nuances of social interactions and enables the recommendation system to provide more personalized and accurate suggestions. Additionally, the concept of trust incorporates a user's social circle's opinions and preferences, making recommendations more robust and comprehensive~\cite{sherchan_nepal_paris_2013}. In this paper, we investigate trust connections, and the implications they have on a user's preferences and the robustness to adversarial attacks.

There are several types of recommendation systems, including content-based, collaborative filtering, hybrid, and others. Content-based recommendation systems suggest items based on the items' characteristics and the users' preferences \cite{haruna2017context}. On the other hand, collaborative filtering recommends items based on the preferences and behaviors of similar users~\cite{su_khoshgoftaar_2009}. Hybrid systems combine both approaches \cite{afoudi2021hybrid}. Trust-based collaborative filtering \cite{davoudi2018social} is particularly interesting because it not only considers the preferences of similar users but also incorporates the trust relationships between users. This approach helps in mitigating some of the limitations of traditional collaborative filtering, such as the cold start problem and data sparsity, by leveraging the trust network to generate recommendations even when there is limited interaction data available~\cite{yang_lei_liu_li_2017}. We will explore this technique in this paper, but with a first principles approach to develop an understanding of how to utilize these different methods and information forms best.

\textbf{Outline.} The paper is organized in the following manner. Section~\ref{lt_re} gives an overview of prior work, and section~\ref{contri} outlines our contribution. After that, in section~\ref{method} we introduce the methodology and in section~\ref{single} we discuss the single information-based algorithms and subsequently multiple information-based algorithms in section~\ref{multiple}. Validation of the findings on more datasets is given in section~\ref{sec:valid}, comaprison against other state-of-the-art (SOTA) algorithms provided in section~\ref{sec:sota} and the impact of adversarial attacks is explored in section~\ref{sec:adversary}. A discussion of our main findings is given in section~\ref{discussion}.

\section{Literature Review} 
\label{lt_re}
\subsection{Traditional Approaches}
Traditional trust-based recommendation systems mainly focus on incorporating explicit and implicit trust relationships into the recommendation process. For example, some approaches consider the ratings or preferences of a user's trusted friends when generating recommendations. Messa et. al \cite{messa_2007} delved into a trust-aware collaborative filtering approach. Their method distinctively pivots on explicit trust relationships. This means that the system directly harnesses expressed trust values (like friendship connections) to modulate the influence of a user's peers on their recommendations.

Other methods incorporate trust propagation mechanisms to infer trust relationships between users who are not directly connected. The idea is grounded in the notion that trust can be transitive to some degree~\cite{richters_peixoto_2011}. For example, if A trusts B and B trusts C, even if A hasn't interacted with C, there might be a certain level of implicit trust A has for C due to the mutual connection. Numerous methods have been developed based on this trust transitivity principle. Some models utilize matrix factorization techniques to propagate trust across the user-item matrix, while others employ graph-based algorithms to trace the spread of trust across a network of users. Such propagation methods attempt to fill the gaps in the trust matrix, offering a more comprehensive and interconnected trust landscape that can be leveraged for more informed and nuanced recommendations~\cite{jamali_ester_2010}.

\subsection{Cold-Start Approaches}

Addressing the ``cold start'' problem remains a pivotal challenge for trust-based recommendation systems. The ``cold start'' issue surfaces when a new user joins the platform or a new item is added, and there's minimal to no interaction data available for meaningful recommendation generation. In such scenarios, traditional collaborative filtering methods often fall short, as they rely predominantly on historical interaction data~\cite{moghaddam2019cold}.  Various solutions have been proposed, including adaptive meta-learning \cite{sanjeevi2015adaptive}, deep learning \cite{wei2017collaborative}, and a demographic approach based on finding similarities between old and new users~\cite{pandey2016resolving}.

Another porpular approach is leveraging the trust network~\cite{lam_vu_le_duong_2008}. The assumption here is that even if a user hasn't interacted with many items, their trusted connections might have. By leveraging this network, systems can extrapolate the new user's potential interests based on their trusted peers' preferences. One particularly noteworthy approach to leverage the trust network against the ``cold start'' problem was conceived by \cite{10.1145/1557019.1557067}. They introduced TrustWalker, a random walk model grounded in the trust network. It commences its journey from the target user and propagates through their trusted connections, and with some probability, swapping to other similar items that may have more information. In this way, TrustWalker harnesses both explicit and implicit trust connections along with skipping to different items, which lets ``cold start'' users with little information still receive ``acceptable'' recommendations.

\subsection{Machine Learning Approaches}

Machine learning, over the past decade, has revolutionized numerous domains with its ability to extract patterns from vast data sets. Traditional methods of collaborative filtering were primarily linear. However, the dynamics of user preferences, trust relationships, and item attributes often exhibit nonlinear patterns that require more sophisticated models for accurate representation \cite{marlin}. Hence, machine learning is a widely used tool in recommendation systems~\cite{khanal2020systematic}. Bayesian~\cite{ericson2013performance, felden2007recommender} and decision tree~\cite{shinde2013scenario, lucas2012making} approaches are mostly used due to their simplicity~\cite{portugal2018use}. Moreover, gradient-decent-based~\cite{chen2013error}, matrix factorization-based~\cite{schelter2013distributed, wang2023improved}, non-linear collaborative filtering \cite{jain2020emucf}, and hidden Markov model-based~\cite{fang2022research} algorithms are also in use. 

Neural collaborative filtering models underscore the potential of deep neural networks in handling the collaborative filtering challenge on implicit feedback data sets~\cite{he_liao_zhang_nie_hu_chua_2017, wang2023collaborative}. They propose three different ML-based models for developing recommendation systems and show the models compete with other state-of-the-art approaches.  Moreover, these approaches have also been used to tackle the cold-start problem~\cite{wei2017collaborative, volkovs2017dropoutnet}.

In parallel, the application of graph neural networks (GNNs) in recommender systems has witnessed a significant surge, establishing GNNs as a state-of-the-art approach in this area. \cite{gnn_survey} offer a comprehensive review of the literature centered around GNN-based recommender systems. The motivation to incorporate GNNs into recommender systems primarily stems from their ability to capture the higher-order structure of the graph . 

\subsection{Adversarial Defense Approaches}

Recommendation systems are vulnerable to adversarial attacks, subverting their performance. These adversarial attacks can manifest in various forms: fake user profiles, strategically crafted reviews, or even swarming behaviors intended to boost or suppress particular items. A robust recommendation system isn't just about accurate recommendations; it's also about ensuring its resilience against these adversarial intrusions~\cite{deldjoo_noia_merra_2021}. Even the most advanced models are vulnerable to certain adversarial tactics~\cite{merra_2021}. However, the silver lining is the evolution of defense mechanisms. Techniques such as adversarial training, where the model is trained with perturbed data, have shown promise in enhancing model robustness.

Low-rank defenses~\cite{entezari2022low} against adversarial attacks are one type of adversarial defense strategy used in recommendation systems. Low-rank defenses entail transforming the user-item matrix into a low-rank matrix to decrease the impact of adversarial attacks. This method has been demonstrated to be effective in fighting against several sorts of adversarial attacks, including poisoning attacks and model extraction attacks. In recommendation systems, generative adversarial networks (GANs) can detect and fight against adversarial attacks \cite{liang2021text}. GANs can be used to generate adversarial instances and teach the recommendation model to defend against them. This technique effectively defends against adversarial assaults such as poisoning and model extraction~\cite{deldjoo_noia_merra_2021}. Moreover, methods based on these have shown great prospects in terms of modeling the long-tail distribution of recommendation systems \cite{qin2024gacrec}. Another study by~\cite{lian_zhao_2022} focused on how to accurately portray the representative rating for an item to mitigate the risk of an adversary adding fake ratings. The paper found that placing weights on network-related factors leads to increased robustness to adversaries.  

\section{Our Contribution} \label{contri}
We aim to leverage trust-based social network data to develop recommendation systems that not only provide accurate recommendations but also can handle the aforementioned challenges around cold-start users and adversarial attacks. We discover that incorporating the trust graph data can facilitate the establishment of recommendation systems that are robust against adversarial attacks and provide high-quality recommendations for cold starters.

This research takes a first principles approach to building a recommendation system. We will develop novel approaches to combining different data forms in recommendation systems, such as a simple, but effective approach for determining item similarity and an adaptable framework for designing recommenders based on user similarities. We build several recommendation systems, starting by only accessing one type of data at a time. This will let us then evaluate these recommendation systems and understand how derivatives of different types of data act in various contexts. 

To structure the development of these recommendation systems, the data can be split into three sub-sets, which are explained and then visualized below. The idea is, that if we can build recommendation systems that are limited to only these fundamental elements of ``information'', we can then take what we learn to produce more effective recommenders that operate on combinations of these subsets.

\begin{itemize}
    \item \textbf{Trust Graph.} The trust graph is simply the nodes and the trust connections between one another. Recommendation systems built upon this sub-set of the data only concern themselves with the structure of the social network and not with the ratings on items.
    \item \textbf{Intra-Item Information.} The intra-item information is a subset of the data that contains the intra-item similarity, essentially just a table, providing similarity scores between pairs of items along with each node's neighbors.
    \item \textbf{Item-Rating Information.} The Item-Rating information is a subset of the data that only has nodes, items, and their ratings. Recommendation systems built upon this subset draw similarities between nodes based on how they've jointly rated items, not considering the actual social connection between them.
\end{itemize}

    \begin{figure}[h!]
    \label{fig-data-type}
    \includegraphics[width=11cm]{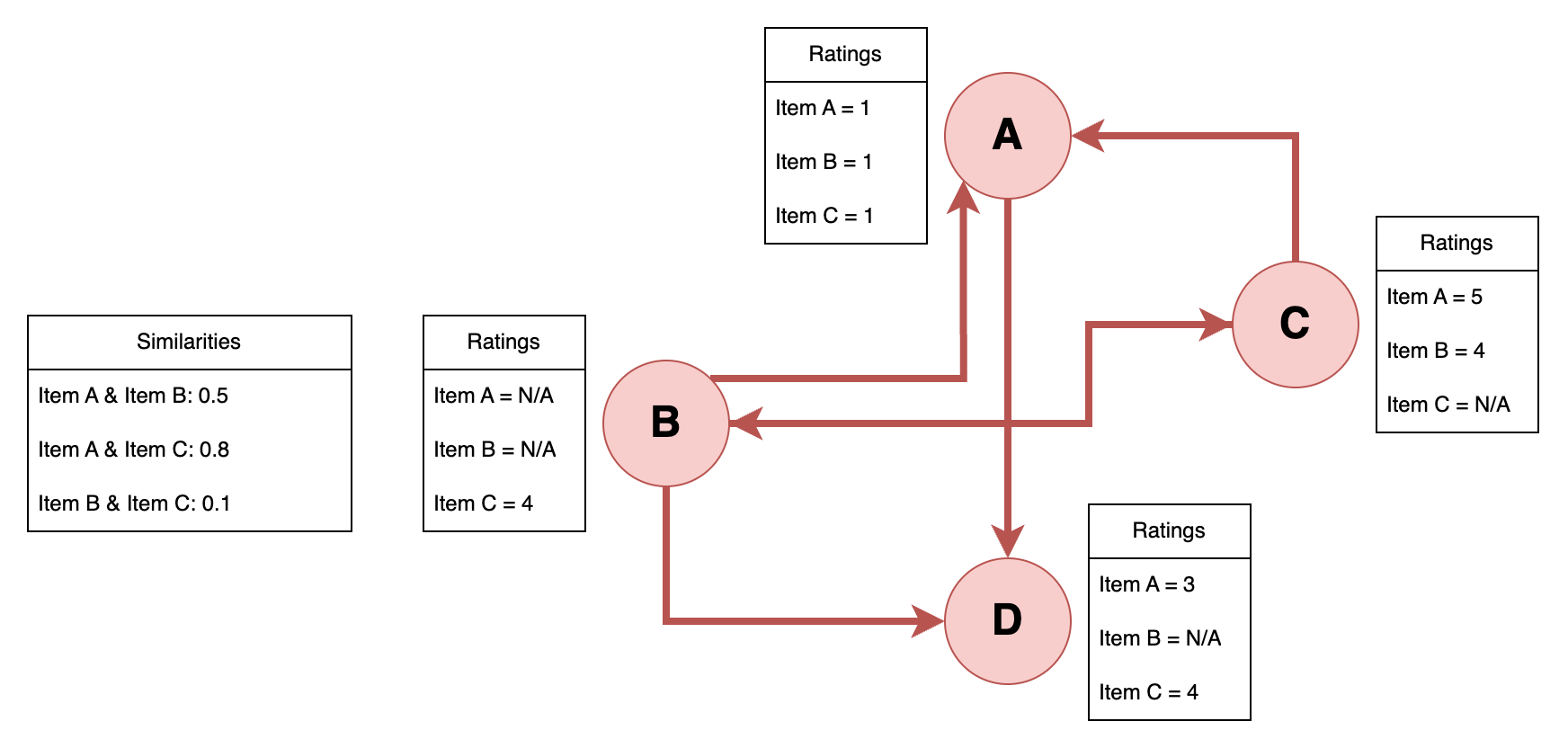}
    \centering
    \caption{An example of three types of information. 4 individuals whose social structure is shown via edges. Rating corresponding to each individual and overall similarity metric among items provided using tables.}
    \end{figure}

\begin{figure}[h!]
    \includegraphics[width=9cm]{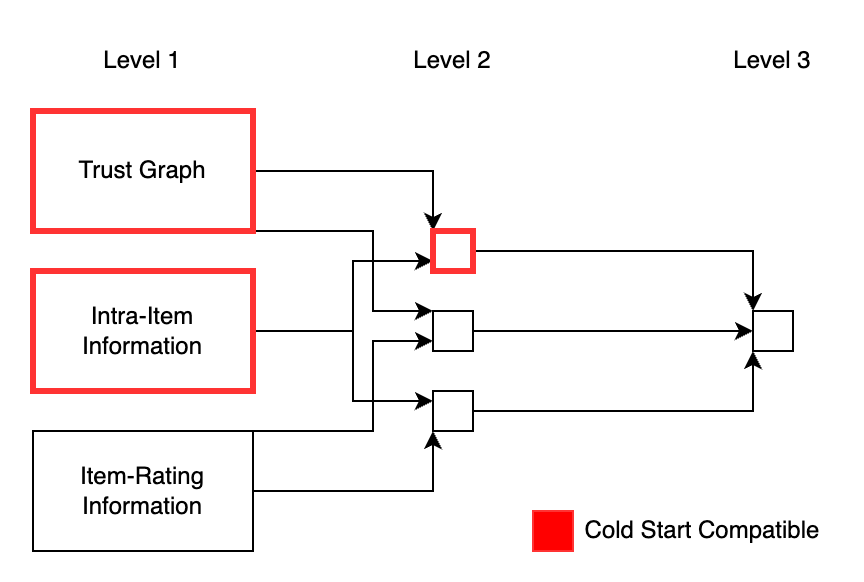}
    \centering
    \caption{Schematic diagram of building a recommendation system based on different subsets of data and combining them. In particular, cold-start compatible approaches are outlined in red. \label{fig-struct}}
\end{figure}

As visualized in Figure~\ref{fig-struct}, we study algorithms in three levels. In level 1, we investigate the algorithms which use one type of information. Building on our findings in level 1, we consider all three possibilities of combining this information in pairs in level 2. Finally, in level 3, we leverage all three types of information to devise algorithms. This systematic study of the problem will allow us to evaluate the strengths of various novel methods, such as similarity indices-based approaches~\cite{bag2019efficient}, random walks techniques~\cite{andersen2008trust}, and opinion formation models~\cite{zehmakan2024majority}. More precisely, we characterize the approaches and types of information that facilitate more accurate performance and robustness against adversarial attacks.

\section{Methodology} \label{method}

\subsection{Primary Training Data Set}

For the sake of iteration in our experiments, we used a single data set for developing and testing the algorithms and then validated the results with other datasets as well.

Epinions data set is one of the most commonly used datasets in building recommendation systems. Epinions was a consumer review site, where users could place reviews on items, and ``trust'' other users. Ratings were based on stars and are hence in the range of 1-5.  The trust network contains 49k users, with 480k ``trust'' edges between them. There are also 150k items, with 660k ratings between them. This data set contains nodes that have at least 1 ``in-edge'', implying at least one other user trusts them. This network is distributed similarly to other social networks, where node popularity exponentially decays, and where there are a few trusted authorities who provide powerful and centralized opinions.

\begin{figure}[h!]
\centering
\includegraphics[width=6cm]{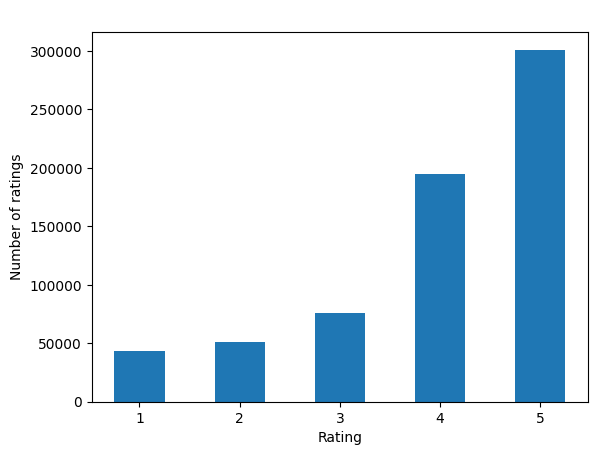}
\caption{Distribution of ratings for all items\label{fig-eopinion}}
\end{figure}

By observing the ratings data (see Figure~\ref{fig-eopinion}), we found that people's ratings are negatively skewed, i.e. users had a positive bias on their ratings relative to the 5-star scale. Upon focusing on the top 10 items rated by the largest number of users, we can observe negative skewness. Hence, being able to accurately provide a recommendation will require interpolation from the graph data and is not easily solved by simply assuming high ratings.

\subsection{Similarity Metrics}
Here, we discuss the similarity metrics used in the design of the recommendation systems.

Note that the trust relations between the users can be modeled using a graph $G=(V, E)$, where the node set $V$ corresponds to the users, and the edge set $E$ models the relations.

\subsubsection{Jaccard Index}
One of the most popular similarity measures between nodes (users) in a graph is the Jaccard index. Let $N(u)$ for a node $u$ denote the set of neighbors of $u$. The Jaccard similarity index between two nodes $u_i$ and $u_j$ is defined as:

$$ J(u_i, u_j) = \frac{|N(u_i)\cap N(u_j)|}{|N(u_i)\cup N(u_j)|} $$

One drawback of this index is that if two nodes are adjacent but share no other neighbors, then the value is 0. To mitigate this issue, we redefine the index in the following way:

$$J'(u_i, u_j) = J(u_i,u_j) + 
\begin{cases}
    \frac{1}{|N(u_i) \cup N(u_j)|}  & u_j \in N(u_i) \\
    0 & \text{otherwise}
\end{cases}$$

\subsubsection{Jaccard-Item Index}
Jaccard-Item Index is a form of the Jaccard index, but instead of relating users using their neighbors, we relate them based on the items they have rated.

$$J_I(u_i, u_j) = \frac{|I(u_i)\cap I(u_j)|}{|I(u_i)\cup I(u_j)|}$$

where $J_I$ is the Jaccard-Item Index and $I(u)$ is the items rated by node $u$. 

\subsubsection{Item-Rating Difference}
The Item-Rating difference compares two nodes as follows:

$$R_D(u_i, u_j) = 1 - \frac{\sum_{k \in I(u_i) \cap I(u_j)}{|R(u_i, k) - R(u_j, k)|}}{4\cdot |I(u_i)\cap I(u_j)|}$$

$R(u_i, k)$ denotes the rating received by the \textit{k}-th item from user $u_i$.  we are effectively enumerating the items that two users have both rated. For each item, we take the absolute difference in their rating. We then divide by 4 to normalize (maximum possible difference in ratings) and divide by the magnitude of the intersection to obtain the average.

\subsubsection{Pearson Correlation}
One similarity metric used to determine the relationship between two items is the Pearson correlation of the ratings given to them. This was proposed by~\cite{10.1145/1557019.1557067}. 

The similarity function between two items, \textit{i} and \textit{j}, works as follows:

$$S(i, j) = \text{set of users who have rated both items } i \text{ and } j.$$

$$\rho(i, j) = \frac{\sum_{u \in S(i, j)}{(R(u, i) - \overline{R}(u))(R(u, j) - \overline{R}(u))}}{\sqrt{(\sum_{u \in S(i, j)}{(R(u, i)-\overline{R}(u))^2})(\sum_{u \in S(i, i)}{(R(u, i)-\overline{R}(u))^2)}}}$$

We then want to skew this correlation, such that if $|S(i, j)|$ is large, we are more confident. So, the final similarity metric, which is equivalent to that given in~\cite{10.1145/1557019.1557067}, is:

$$sim(i, j) = \frac{1}{1 + e^{-\frac{|S(i, j)|}{2}}} \times \rho(i, j)$$

The sigmoid function is used to balance how much we favor the magnitude of $S$ and to ensure our similarity metric is in the range $[0, 1]$.

Thus, for any pair of items, using just the Intra-Item graph, we can determine a quantified value that indicates how ``related'' the pair of items is. 

\subsubsection{Intra-Item Jaccard}
In this similarity metric, to relate items we use Jaccard. i.e. the number of users who have rated $i$ and $j$ divided by the union of the users who rated both $i, j$. So the similarity between the two items is:

$$J_{II}(i, j) = \frac{|S(i,j)|}{|S(i) \cup S(j)|}$$

where $S(i)$ and $S(j)$ are the set of nodes rated $i$ and $j$, respectively.
\subsection{Evaluating Recommendation Systems}

The information needed by recommendation systems is quite sparse. Considering the number of items and graph size and often ratings concentrated on only popular items, computing recommendations for every item, is very costly. Hence, we will be focusing on an approach that will give results relatively quickly. To make recommendations quickly, and to build an environment where there is substantive data, we reduce the data sets to a subset where we keep the top $k$ items and nodes that have rated these items. Now the subset contains active users, who all have at least one item in common (the most popular item), and it is likely that they also share a number of the other top items due to their popularity. 

To quantify the performance, we take $15\%$ of the users at random, who are called ``test users'' and now for each item $i$, we remove its true rating and then compute the recommendation for the item. After that, we calculate the binary accuracy (correct discrete rating or not) and mean absolute error (MAE)~\cite{bobadilla2013recommender} across all the test users.  This process is repeated 5 times and for each time, we record the MAE (this is important since some of our algorithms are randomized). For simplicity, we assume the recommender's performance follows a normal distribution. We compute the standard deviation and mean of MAE to gain an idea of how reliable the recommender is. 

As a reference point for designed recommenders, we use two simple strategies as control recommenders. 

\begin{enumerate}
    \item \textbf{Random}: The random recommender simply provides a random rating of 1-5 for each item.
    \item \textbf{Universal Random}: The universal random recommender takes the distribution of ratings into account by generating a rating proportional to the ratings given to the item.
    
\end{enumerate}

Both recommenders on the Epinions data set,  give moderate results and will be good reference points for future evaluation.

\begin{table}[h]
\centering
\caption{Mean and Standard Deviation of MAE for Baseline Algorithms}
\begin{tabular}{lcc}
\toprule
Algorithm & $\mu_{MAE}$ & $\sigma_{MAE}$ \\
\midrule
Random & 1.67 & \textbf{0.21} \\
Universal Random & \textbf{1.30} & 0.38 \\
\end{tabular}
\label{tab:results1}
\end{table}

\begin{figure}[h!]
    \includegraphics[width=7cm]{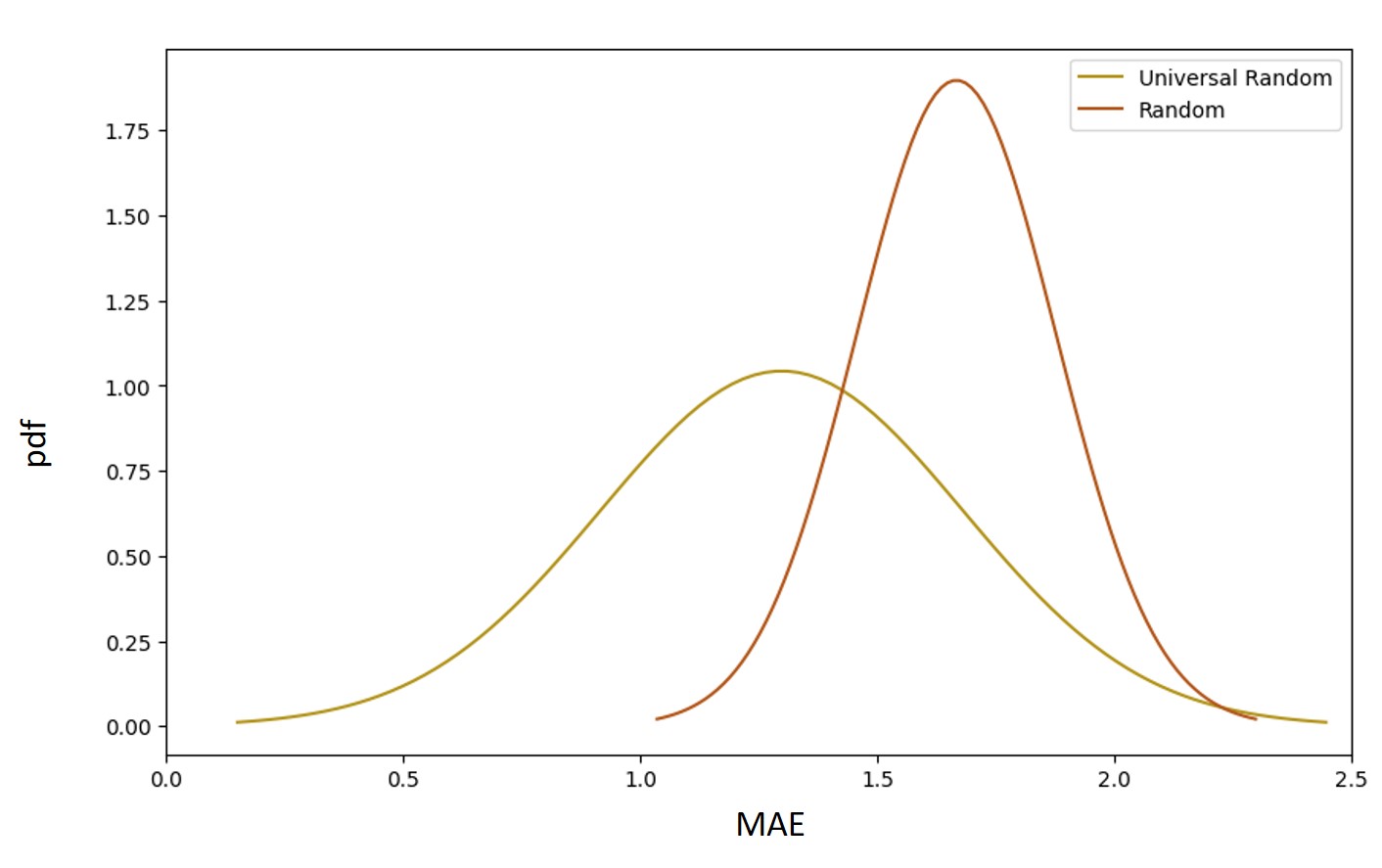}
    \centering
    \caption{Distribution of MAE for Control Recommenders}
\end{figure}

\section{Understanding single level information} \label{single}.

In this section, we focus on the algorithms which use only one of the three types of information available.
\subsection{Trust Graph}

The most simple approach to take with the trust graph is, given a node, to investigate its neighbors. We can infer the current node's rating by measuring its neighbors' central tendencies like mean, median, and mode.

\subsubsection{Centrality of Neighbours}
Let's say we are rating item \textit{i}, we go through a node's neighbors and calculate the mean, median, or mode of their ratings for item \textit{i} (if it exists). If no ratings for \textit{i} exist, we return a random rating. Note, this algorithm mutates the underlying data source, whereby a recommended value for a node, can be used by another node in its rating inference. The performance of these recommenders can be seen in Table~\ref{tab:results2}. We observe that the mean and median of neighbors were the best algorithms in this range. The mode version did not perform as well. This is because mode cannot produce decimal-precision ratings it also defaults to randomness when a mode can't be determined.

\begin{table}[h!]
\centering
\caption{Mean and Standard Deviation of MAE for Neighbourhood Algorithms}
\begin{tabular}{lcc}
\toprule
Algorithm & $\mu_{MAE}$ & $\sigma_{MAE}$ \\
\midrule
Mean of Neighbours & \textbf{1.27} & 0.25 \\
Median of Neighbours & \textbf{1.27} & \textbf{0.19} \\
Mode of Neighbours & 1.32 & 0.23 \\
Universal Random & 1.30 & 0.38 \\
Random & 1.67 & 0.21 \\
\end{tabular}
\label{tab:results2}
\end{table}

\subsubsection{Jaccard-Weighted Model}
The Jaccard-Weighted Model extends the neighbor-based models from before. Effectively, given the neighbors of node $u$, their rating of an item is considered based on their modified Jaccard Index ($J'$) with node $u$. Finally, we identify the rating that received the maximum `score'. The performance of this algorithm, along with the previous Median of Neighbours model can be seen in Table~\ref{tab:results3}. We observe that the Jaccard-Weighted model performs quite similarly to the Median of Neighbours model when it comes to mean accuracy.

% However, the distribution is wider and the peak lower, implying the algorithm is more noisy.

\begin{table}[h!]
\centering
\caption{Mean and Standard Deviation of MAE for Jaccard Weighted Neighbours}
\begin{tabular}{lcc}
\toprule
Algorithm & $\mu_{MAE}$ & $\sigma_{MAE}$ \\
\midrule
Median of Neighbours & \textbf{1.27} & \textbf{0.19} \\
Jaccard Weighted Neighbours & 1.31 & 0.32 \\
Universal Random & 1.30 & 0.38 \\
Random & 1.67 & 0.21 \\
\end{tabular}
\label{tab:results3}
\end{table} 

\subsubsection{Monte-Carlo Random Walk}

We now move to investigate recommendation systems that consider the entire trust graph. The Monte-Carlo (MC) Random Walk model works by starting at the node we are trying to produce a rating for and then performing a random walk. As the distance from the original node, $d$, increases, so does the probability that a random rating is returned. This random walk continues until a node with a rating is reached, or this random rating is produced. We run this experiment $k$ times, for some fixed $k$, for each node and take the mean rating from the nodes reached via random walk. Here $\alpha$ controls probability of making a random jump to a rated node instead of continuing the regular traversal. The pseudocode for this algorithm is given in Algorithm~\ref{alg:majority}.

\begin{algorithm}[H]
\caption{Monte Carlo Random Walk}\label{alg:majority}
\begin{algorithmic}[1]
\Function{monteCarloRandomWalk}{$node$, $item$, $graph$, $ratings$, $k$}

    \Function{traverse}{$node$, $distance=0$, $\alpha=0.05$}
        \If{$node \textbf{ in } ratings$}
            \State \textbf{return} $ratings[node]$
        \EndIf

        \If{$\text{random(0, 1)} < d \times \alpha$}
            \State \textbf{return} $traverse(random.choice(ratings)])$
        \EndIf
        
        \State $next \gets graph[node]$
        \If{$|next.neighbours| == 0$}
            \State \textbf{return} $\text{None}$
        \EndIf
        
        \State \textbf{return} $traverse(random.choice(next.neighbours)), d + 1)$
    \EndFunction

    \State $results \gets []$
    \For{$i \textbf{ in } \text{range}(k)$}
        \State $sampleRating \gets traverse(node)$
        \If{$sampleRating == None$}
            \State $results.push(random(1,5))$
        \Else
            \State $results.push(sampleRating)$
        \EndIf
    \EndFor

\State \textbf{return} $mean(ratings)$
\EndFunction
\end{algorithmic}
\end{algorithm}

We then extended this algorithm by choosing the neighbor, at each step of the random walk based on a weighted random incorporating the Jaccard index. The performance of both MC algorithms, as compared with the others can be seen in Table~\ref{tab:results4}. We observe that the Monte Carlo algorithm outperforms the previous models. Interestingly, the Jaccard weighted version performed better. This implies that Jaccard indexes seem to hold more valuable information than mere trust connections.

\begin{table}[h!]
\centering
\caption{Mean and Standard Deviation of MAE for Random Walk Algorithms}
\begin{tabular}{lcc}
\toprule
Algorithm & $\mu_{MAE}$ & $\sigma_{MAE}$ \\
\midrule
Jaccard Monte-Carlo Random Walk & \textbf{1.14} & 0.22 \\
Monte-Carlo Random Walk & 1.16 & 0.23 \\
Median of Neighbours & 1.27 & \textbf{0.19} \\
Universal Random & 1.30 & 0.38 \\
Random & 1.67 & 0.21 \\
\end{tabular}
\label{tab:results4}
\end{table}

% \begin{figure}[h!]
%     \includegraphics[width=10cm]{MC_random_walks.jpg}
%     \centering
%     \caption{Distribution of MAE for Random Walk-based models}
% \end{figure}

\subsubsection{Jaccard Majority of Majorities (JMoM) and Weighted Average (Jaccard WA)}
These recommenders work by considering every other node, which has rated item \textit{i}, in the graph. We then perform a summation of the Jaccard values for every node corresponding to each rating value. In the Majority of Majorities version, we then take the rating that has received the highest contribution of Jaccard scores. However, in the Weighted Average version, we instead take the weighted average of these ratings, weighted by the contribution of Jaccard scores. We will call these techniques JMoM and WA moving forward. (These methods are inspired by the popular majority based opinion diffusion models, cf.~\cite{gartner2018majority,auletta2020complexity,zehmakan2019spread}.). The pseudocode for weighted average is given in Algorithm~\ref{alg:WA}.

\begin{algorithm}[H]
\caption{Weighted-Average}\label{alg:WA}
\begin{algorithmic}[1]
\Function{weightedAverage}{$node$, $item$, $graph$, $ratings$, $similarityFunction$}

\State $buckets \gets Array(5)$
\For{$otherVertex \textbf{ in } graph$}
    \If{$otherVertex \textbf{ in } ratings \textbf{ \&\& } otherVertex \neq node$}
        \State $idx \gets ratings[otherVertex][item] - 1$
        \State $buckets[idx] \gets buckets[idx] + similarityFunction(current, otherVertex)$
    \EndIf
\EndFor
\If{$potentialRatings.sum == 0$}
    \State \textbf{return} $ratings$ randomInteger(1, 5)
\EndIf

\State $WA \gets 0$

\For{$i \textbf{ in } range(5)$}
    \State $WA \gets WA + buckets[i]*(i+1)$
\EndFor

\State \textbf{return} $WA$
\EndFunction
\end{algorithmic}
\end{algorithm}

As can be observed from Table~\ref{tab:results5}, the JMoM performed reasonably well. Its mean accuracy is better than the random walk. However, it was less robust. The WA version performed very well significantly beating both the JMoM version and the random walk. By overcoming the constraint of only considering direct neighbors, the JMoM and Jaccard WA models can produce decent results, with a very simple algorithm. Note that these approaches can be used with any user-to-user similarity metric.

\begin{table}[h!]
\centering
\caption{Mean and Standard Deviation of MAE for Full-graph Jaccard Algorithms}
\begin{tabular}{lcc}
\toprule
Algorithm & $\mu_{MAE}$ & $\sigma_{MAE}$ \\
\midrule
Jaccard WA & \textbf{1.05} & 0.25 \\
Jaccard Monte-Carlo Random Walk & 1.14 & 0.22 \\
Jaccard MoM & 1.13 & 0.30 \\
Universal Random & 1.30 & 0.38 \\
Random & 1.67 & \textbf{0.21} \\
\end{tabular}
\label{tab:results5}
\end{table}

% \begin{figure}[h!]
%     \includegraphics[width=10cm]{JMM_performance.jpg}
%     \centering
%     \caption{Distribution of MAE for Jaccard Index-based models}
% \end{figure}

\subsection{Item Rating Information}

In the Item Jaccard model, we used the Jaccard-Item Index $J_I$ which is based on items rated. We also used the weighted average framework approach but with the Item-Rating Difference. The performance of these alternative WA models can be seen in Table~\ref{tab:results6}.

\begin{table}[h!]
\centering
\caption{Mean and Standard Deviation of MAE for Item Rating Algorithms}
\begin{tabular}{lcc}
\toprule
Algorithm & $\mu_{MAE}$ & $\sigma_{MAE}$ \\
\midrule
Item-Jaccard WA & \textbf{1.02} & 0.23 \\
Jaccard WA & 1.05 & 0.25 \\
Item-Rating Difference WA & 1.17 & 0.49 \\
Universal Random & 1.30 & 0.38 \\
Random & 1.67 & \textbf{0.21} \\
\end{tabular}
\label{tab:results6}
\end{table}

% \begin{figure}[h!]
%     \includegraphics[width=10cm]{item-based-wa-performance.jpg}
%     \centering
%     \caption{Distribution of MAE for Item Rating based models}
% \end{figure}

We can see that the Item-Jaccard form of WA outperforms the previous Jaccard WA. Item-Rating Difference doesn't seem to perform very well and is quite inconsistent in its performance. A potential reason for the Item-Jaccard WA model performing so well is that a user's social circle may not be the best predictor for who is similar to them, rather relating people by their items themselves does a better job.

\subsection{Intra-Item Information}
The Intra-Item Information concerns itself with the relationships between items themselves. In a similar fashion to the previous algorithms, we manipulated the WA approach to work with Intra-Item similarity data. This is slightly different from the WA framework as we iterate through each user's items rather than using a user-user similarity metric. The performance of the algorithms, one based on $J_{II}$ and one based on the Pearson $sim$, can be seen in Table~\ref{tab:results7}.

\begin{table}[h!]
\centering
\caption{Mean and Standard Deviation of MAE for Intra-Item Algorithms}
\begin{tabular}{lcc}
\toprule
Algorithm & $\mu_{MAE}$ & $\sigma_{MAE}$ \\
\midrule
Item-Jaccard WA & \textbf{1.02} & 0.23 \\
Intra-Item WA & 1.24 & 0.14 \\
Universal Random & 1.30 & 0.38 \\
Intra-Item WA (Pearson) & 1.34 & \textbf{0.13} \\
Random & 1.67 & 0.21 \\
\end{tabular}
\label{tab:results7}
\end{table}

% \begin{figure}[h!]
%     \includegraphics[width=10cm]{intra-item-sim-performance.jpg}
%     \centering
%     \caption{Distribution of MAE for Intra-item based models}
% \end{figure}

It is evident both of the intra-item-based models underperform our previous models. The intra-item similarity using the Jaccard approach outperforms the Pearson correlation approach. This is a very interesting result, potentially this could be a better metric for determining the similarity of items and could be applied in~\cite{10.1145/1557019.1557067}. It is evident the intra-item similarity conveys some information as we perform better than random, hence this could be useful in combination models in the coming sections.

\subsection{Summary}
By splitting the data up in this way, we now understand how algorithms focused on the different types of information perform. In a way, we know the value of the different types of information. Not very surprisingly, our item-based similarity metrics resulted in better-performing algorithms. However, as we will see in the rest of the paper, leveraging the other types of data facilitate the design of algorithms which are more robust against adversarial attacks and support cold start users.

\section{Combining Multiple Information} \label{multiple}

\subsection{Jaccard Intra-Item WA}
As previously discussed, both Trust and Intra-Item algorithms are effective on cold start users, and so their combination. In an attempt to combine both these modes of data, we developed an intersection between the Intra-Item WA ($J_{II}$) which operates on neighbors, and the Jaccard WA which operates on the entire graph. The idea of this algorithm is for every (user, item) pair, we contribute the Jaccard index of the user multiplied by the item's similarity to our rating bins. The performance can be seen in Table~\ref{tab:results8}.

\begin{table}[h!]
\centering
\caption{Mean and Standard Deviation of MAE for Jaccard Intra-Item WA Algorithm}
\begin{tabular}{lcc}
\toprule
Algorithm & $\mu_{MAE}$ & $\sigma_{MAE}$ \\
\midrule
Item-Jaccard WA & \textbf{1.02} & 0.23 \\
Jaccard Intra-Item WA & 1.20 & \textbf{0.13} \\
Universal Random & 1.30 & 0.38 \\
Random & 1.67 & 0.21 \\
\end{tabular}
\label{tab:results8}
\end{table}

% \begin{figure}[h!]
%     \includegraphics[width=10cm]{jaccard-intra-item.jpg}
%     \centering
%     \caption{Distribution of MAE for Jaccard Intra-Item WA models}
% \end{figure}

Combining these two forms of information seem to produce a very consistently performing recommender. However, it did not perform very well relative to our current best, Item-Jaccard WA in terms of mean MAE.

\subsection{Weighted Item-Rating Difference WA (WIRD WA)}
This model combines the Intra-Item graph with the Item graph. Thus combing information about how nodes rated individual items and how the items are interrelated. The general idea was to slightly alter the item rating similarity metric, $R_D$, to provide a bias taking into account the similarity of items to the item we are recommending.

$$WR_D(u_i, u_j, k) = 1 - \frac{\sum_{j \in I(u_j)}{J_{II}(k, j) \times |R(u_i, k) - R(u_j, j)|}}{4\cdot |I(u_j)|}$$

Note for intra-item similarity we use $J_{II}$ rather than $sim$ as we found it was more performant in the previous section.
The results can be seen in Table~\ref{tab:results9}:

\begin{table}[h!]
\centering
\caption{Mean and Standard Deviation of MAE for WIRD WA Algorithm}
\begin{tabular}{lcc}
\toprule
Algorithm & $\mu_{MAE}$ & $\sigma_{MAE}$ \\
\midrule
WIRD WA & 1.05 & 0.28 \\
Item-Jaccard WA & \textbf{1.02} & 0.23 \\
Intra-Item WA & 1.34 & \textbf{0.13} \\
Item-Rating Difference WA & 1.17 & 0.49 \\
Universal Random & 1.30 & 0.38 \\
Random & 1.67 & 0.21 \\
\end{tabular}
\label{tab:results9}
\end{table}

% \begin{figure}[h!]
%     \includegraphics[width=10cm]{WIRD.jpg}
%     \centering
%     \caption{Distribution of MAE for WIRD WA models}
% \end{figure}

Though the WIRD model performed quite well still not better than the Item-Jaccard WA model. However, it performs better than just the intra-item information or item-rating difference WA alone. Hence, we have successfully combined the intra-item and item-rating information to achieve a better result.

\subsection{Jaccard Item-Jaccard WA}
In this model, we combine the Trust information with the Item-Rating information. To do so, we explored different ways of combining the Jaccard WA and Item-Jaccard WA models. Both similarity metrics are in the range $[0, 1]$, so we attempted linear combinations, multiplying them and using the $max$ function.

Thus, the metrics would be combined as:
$$I_S(u_i, u_j) = \alpha \times J'(u_i, u_j) + (1-\alpha) \times J_I(u_i, u_j), \alpha \in [0, 1]$$

\begin{table}[h!]
\centering
\caption{Mean and Standard Deviation of MAE for Jaccard Item-Jaccard WA Algorithm}
\begin{tabular}{lcc}
\toprule
Algorithm & $\mu_{MAE}$ & $\sigma_{MAE}$ \\
\midrule
Jaccard Item-Jaccard WA (Addition, $\alpha=0.5$) & \textbf{1.00} & 0.26 \\
Jaccard Item-Jaccard WA (Maximum) & 1.01 & 0.26 \\
Jaccard Item-Jaccard WA (Multiplication) & 1.04 & 0.22 \\
Item-Jaccard WA & 1.02 & 0.23 \\
Universal Random & 1.30 & 0.38 \\
Random & 1.67 & \textbf{0.21} \\
\end{tabular}
\label{tab:results11}
\end{table}

% \begin{figure}[h!]
%     \includegraphics[width=10cm]{jaccard-item-jaccard.jpg}
%     \centering
%     \caption{Distribution of MAE for Jaccard Item-Jaccard WA models}
% \end{figure}

Looking at the multiplication-based model, it produced an interesting result which was mirrored in the Jaccard and $J_{II}$ combined from the Trust and Intra-Item section. The accuracy was essentially a blend between the Jaccard model and the Item-Difference one. The $max$-based model performed well but not as well as addition. This implies that even small values of Jaccard or Item-Jaccard contribute information to the recommender that the addition model captures. So, we find that the Jaccard Item-Jaccard WA combination model with equal contribution is the best performer.

\subsection{All Combined Algorithm}

Now, we will combine all the data we have in an attempt to build our best-performing model. From previous experiments, we found that the Jaccard cross Item-Jaccard Weighted Average model performed the best. This incorporates the trust information and the rating information. How can we extend this to incorporate intra-item information? To do this, we defined the similarity between two given users as $U_{sim} = J(u_1, u_2) + I_j(u_1, u_2)$. Let's say we are predicting a rating for $u_1$, item $I$. We would iterate through all other nodes in the graph, and for each of the items they've rated, contribute $U_{sim} \times J_{II}(I, j)$ where $j$ is the other item. Essentially, other users influence our user's ratings by how similar they are based on the social network structure, how similar they are based on the items they've rated, and how similar each of their items is to the one we are predicting.
% \begin{figure}[h!]
%     \includegraphics[width=10cm]{all_combined.jpg}
%     \centering
%     \caption{Distribution of MAE for full information models}
% \end{figure}
Interestingly, the combination model under-performed our best model in terms of mean MAE. 

\section{Validation With Other Data Sets}
\label{sec:valid}
Throughout this paper, we have designed our models and tested them using the Epinions data set. This is a data set popular in literature and gives us a good playground. However, to test the universality of the algorithms, we have tested them on the following two extra datasets.

\begin{enumerate}
    \item FilmTrust~\cite{guo2013novel}
    
        It was a movie review platform, where users could `trust' one another's reviews. This data set was scraped from the site in 2011 and is of common use for recommendation system research.
        
    \item CiaoDVD~\cite{guo2014etaf}
    
        This is again another movie review website, with a similar structure to FilmTrust. This data set was scraped in 2013.
\end{enumerate}

\begin{table}
\centering
\caption{Mean and Standard Deviation of MAE for All Algorithms on Different Data Sets}
\begin{tabular}{lccc|ccc|ccc}
\toprule
\multirow{2}{*}{Algorithm} & \multicolumn{3}{c}{Epinions} & \multicolumn{3}{c}{FilmTrust} & \multicolumn{3}{c}{CiaoDVD} \\
& $\mu_{MAE}$ & $\sigma_{MAE}$ & & $\mu_{MAE}$ & $\sigma_{MAE}$ & & $\mu_{MAE}$ & $\sigma_{MAE}$ \\
\midrule
Jaccard Item-Jaccard WA & \textbf{1.00} & 0.26 & & \textbf{0.66} & 0.08 & & \textbf{0.56} & \textbf{0.28} \\
Item-Jaccard WA & 1.02 & 0.23 & & 0.67 & 0.08 & & 0.58 & 0.32 \\
Jaccard WA & 1.05 & 0.25 & & 1.14 & 0.08 & & 1.73 & 0.36 \\
WIRD WA & 1.05 & 0.28 & & 0.67 & \textbf{0.04} & & 0.72 & 0.37 \\
Jaccard Item-Jaccard JII Combination WA & 1.07 & 0.22 & & 0.69 & 0.06 & & 0.63 & 0.30 \\
JWIRD WA & 1.09 & 0.27 & & 0.67 & 0.07 & & 0.72 & 0.40 \\
Jaccard MoM & 1.13 & 0.30 & & 1.19 & 0.09 & & 1.77 & 0.37 \\
Jaccard Monte-Carlo Random Walk & 1.14 & 0.22 & & 1.20 & 0.08 & & 1.81 & 0.49 \\
Monte-Carlo Random Walk & 1.16 & 0.23 & & 1.20 & 0.08 & & 1.82 & 0.35 \\
Item-Rating Difference WA & 1.17 & 0.49 & & 0.67 & 0.08 & & 0.79 & 0.41 \\
Jaccard Intra-Item WA & 1.20 & \textbf{0.13} & & 1.07 & 0.06 & & 1.75 & 0.32 \\
Intra-Item WA & 1.24 & 0.14 & & 1.18 & 0.08 & & 1.70 & 0.45 \\
Median of Neighbours & 1.27 & 0.19 & & 1.26 & 0.08 & & 1.76 & \textbf{0.28} \\
Mean of Neighbours & 1.27 & 0.25 & & 1.25 & 0.07 & & 1.67 & 0.43 \\
Universal Random & 1.30 & 0.38 & & 0.89 & 0.11 & & 0.72 & 0.58 \\
Jaccard Weighted Neighbours & 1.31 & 0.32 & & 1.22 & 0.10 & & 1.66 & 0.44 \\
Mode of Neighbours & 1.32 & 0.23 & & 1.23 & 0.09 & & 1.67 & 0.43 \\
Intra-Item WA (Pearson) & 1.34 & \textbf{0.13} & & 1.21 & 0.07 & & 1.67 & 0.50 \\
Random & 1.67 & 0.21 & & 1.34 & 0.07 & & 1.82 & 0.31 \\
\bottomrule
\end{tabular}
\label{tab:merged_and_sorted_results}
\end{table}

It appears the Jaccard Item-Jaccard WA recommendation algorithm performed best across all data sets with the lowest MAE value. The algorithm consistently outperformed other algorithms in each of the data sets - Epinions, FilmTrust, and CiaoDVD - with MAEs of 1.00, 0.66, and 0.56 respectively.  This algorithm's Mean MAE ($\mu_{MAE}$) was significantly lower than other algorithms.

Several observations can be made from the results in Table~\ref{tab:merged_and_sorted_results}:

\begin{itemize}
\item As expected, the Random algorithm had the highest MAE in every experiment.
\item The performance of algorithms varied across data sets. For instance, the Jaccard WA algorithm had a relatively low MAE for Epinions but performed poorly on the FilmTrust and CiaoDVD data sets. 
\item Algorithms that combined different recommendation techniques, like Jaccard Item-Jaccard JII Combination WA, did not necessarily guarantee better performance than their simpler counterparts.
\end{itemize}

\section{Comparison Against SOTA Algorithms}
\label{sec:sota}

Here, we compare our results with other previously established methods.  We compared against algorithms like SoRec\cite{ma2008sorec}, SoReg \cite{ma2011recommender}, TrustSVD\cite{guo_2015}, TrustMF \cite{yang2016social}, and APURTJ \cite{abdi2025improved}. APURTJ stands for Adjusted
Triangle, User Rating Preferences behavior and Jaccard
similarity. This algorithm outperforms the other methods. 

The optimal experimental settings for each method are suggested by previous works \cite{taheri2017extracting}. The settings are: (1) SoRec: the number of latent
features $d = 5$ for SoRec and $\lambda_c = 1.0, 0.01, 1.0$ corresponding to FilmTrust, CiaoDVD and Epinions respectively; (2) SoReg: the number of latent features d = 5 for SoReg and $\beta = 0.1$ for the all; (3) TrustMF: : the number of latent features $d = 5$ and $\lambda_t = 1$ ; (4) TrustSVD: the number of latent features
$d = 5$ for TrustSVD and $\lambda = 0.1$, $\lambda_t = 0.9$ for FilmTrust, $\lambda = 0.6$, $\lambda_t = 0.5$ for Epinions, and $\lambda = 0.5$, $\lambda_t = 1.0$ for CiaoDVD.

Based on the results given in Table~\ref{tab:sota_results}, we observe that while our algorithms do not necessarily outperform all other methods, their performance is satisfactory, especially considering their simplicity and scalability. It is worth to emphasize that the main goal of this paper is a systematic study of the impact of various sources of information on the recommendation quality, rather than providing a better recommendation algorithm. However, the above comparison has been conducted to complement the understanding of the reader.
\begin{table}[h]
\centering
\caption{Mean of MAE for SOTA Algorithms on Different Data Sets}
\begin{tabular}{lccc}
\toprule
Algorithm & Epinions & FilmTrust & CiaoDVD \\
\midrule
Jaccard Item-Jaccard WA & 1.00 & 0.66 & 0.56 \\
SoReg & 0.96 & 0.64 & 0.9 \\
SoRec & 0.92 & 0.72 & 0.81 \\
TrustSVD & \textbf{0.82}& 0.61 & 0.72 \\
TrustMF & 0.83 & 0.63  & 0.76 \\
ATURPJ & - & \textbf{0.5} &  \textbf{0.55} \\
Universal Random & 1.30 & 0.89 & 0.72 \\
Random & 1.67 & 1.34 & 1.82 \\
\bottomrule
\end{tabular}
\label{tab:sota_results}
\end{table}

\section{Robustness Against Adversarial Attacks}
\label{sec:adversary}

We will now investigate how our recommendation systems respond to adversarial attacks on the network. To conduct this experiment we must construct an attack method and mutate the graph to represent it. The attack we will simulate is that of an actor who wants to artificially reduce the ratings of certain items on the system. The actor can create fake ``bot'' accounts and befriend an adversarial celebrity. We choose the 10th most popular node by in-edges as a celebrity node and then create some number of fake nodes, who rate every item a $1$ out of $5$. These nodes will trust the celebrity node and will be trusted by the celebrity too. 

To test the robustness of our algorithms, we will use the ``Epinions'' data set and use 150 fake accounts (around 5\% of the graph by node size). We deem this to be quite a significant attack on the integrity of the network. We only tested the algorithms that performed reasonably well. Algorithms tested here are as follows: Jaccard Item-Jaccard WA, Item-Jaccard WA, Jaccard WA, WIRD WA, Jaccard Item-Jaccard JII Combination WA, and Jaccard Monte-Carlo Random Walk.

\begin{table}[h!]
\centering
\caption{Algorithm Performance Metrics Under Adversarial Conditions}
\begin{tabular}{lcccc|c}
\toprule
Algorithm & \multicolumn{2}{c}{Normal} & \multicolumn{2}{c}{Adversarial} & \multicolumn{1}{c}{Reduction (\%)} \\
\cmidrule(lr){2-3} \cmidrule(lr){4-5} \cmidrule(lr){6-6}
& $\mu_{MAE}$ & $\sigma_{MAE}$ & $\mu_{MAE}$ & $\sigma_{MAE}$ & $\mu_{MAE}$ \\
\midrule
Jaccard Item-Jaccard WA & \textbf{1.00} & 0.26 & 1.25 & \textbf{0.19} &  24.93 \\
Item-Jaccard WA & 1.02 & 0.23 & 1.24 & 0.20 &  21.77 \\
Jaccard WA & 1.05 & 0.25 & \textbf{1.17} & 0.22 &  12.12 \\
WIRD WA & 1.05 & 0.28 & 1.32 & 0.23 &  25.53 \\
Jaccard Item-Jaccard JII Combination WA & 1.07 & \textbf{0.22} & 1.23 & 0.20 &  15.58 \\
Jaccard Monte-Carlo Random Walk & 1.14 & 0.22 & 1.21 & \textbf{0.19} &  \textbf{6.39} \\
\bottomrule
\end{tabular}
\label{tab:results10}
\end{table}

\begin{figure}[h!]
    \centering 
    \includegraphics[width=11cm]{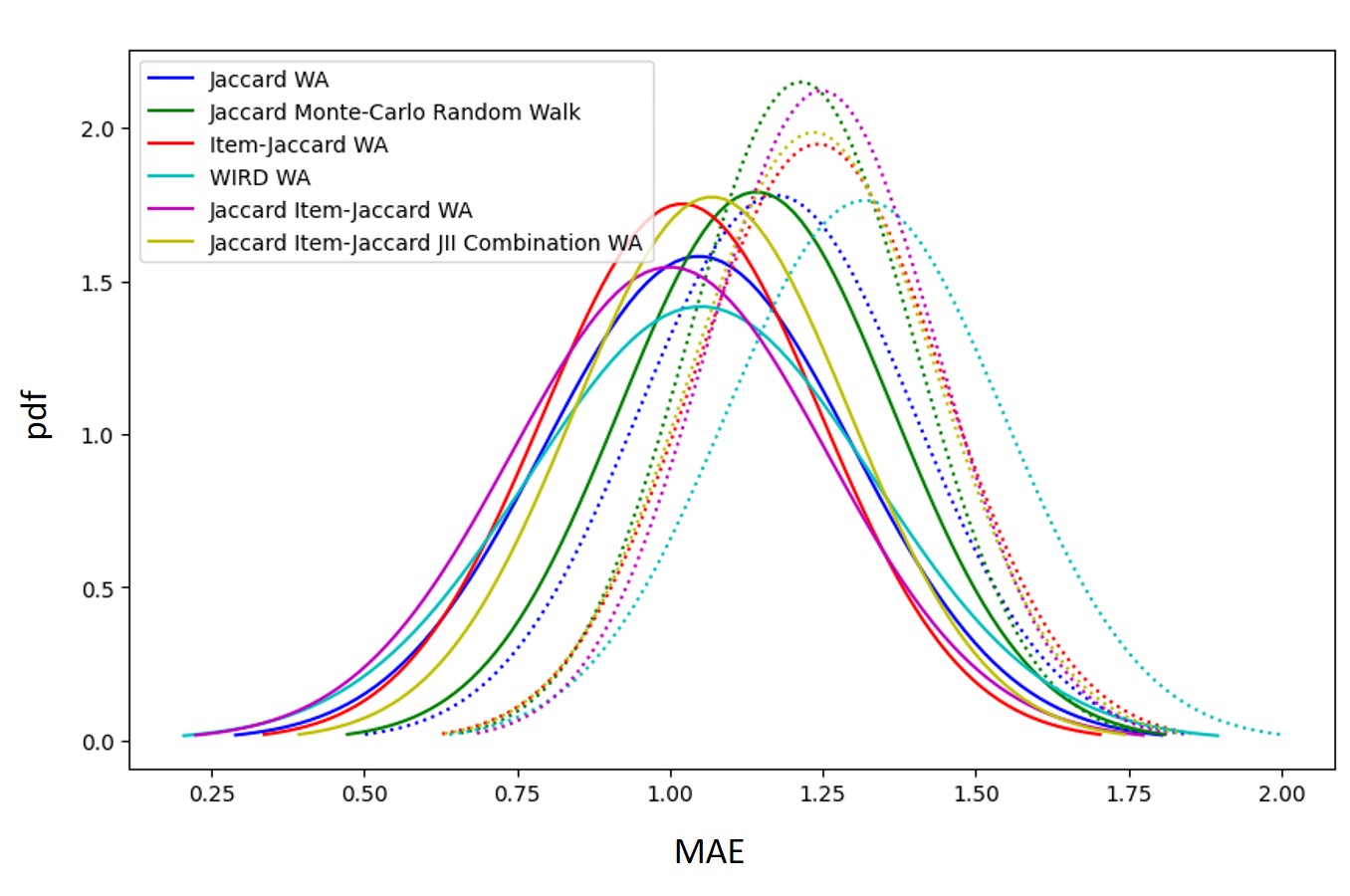}
    \caption{ MAE distributions for different recommender algorithms. The dotted lines depict the performance with an adversarial network and the solid lines show the same recommender under normal conditions.}
    \label{fig:adversary}
\end{figure}

In these results, we can observe that the minimally impacted algorithms were those that relied on the trust information. These are the Jaccard WA model and the Jaccard Monte-Carlo Random Walk. The most impacted were those that relied on Item-Rating and Intra-Item information, i.e. Item-Jaccard WA and WIRD WA. Interestingly, the combination model, which made use of all forms of data, seemed to not suffer as big a reduction in performance.

\section{Discussion} \label{discussion}

We successfully broke down the concept of a recommender system to its core principles, starting with the information it ingests. Through breaking down the different information forms, we noticed that the most important feature when assigning the similarity between two users for their preferences is the way they've rated other items and which items they have engaged with. The trust and intra-item-based algorithms led to recommenders with higher MAE as seen in Table~\ref{tab:merged_and_sorted_results}.  All other models included the item-rating information in some way, which had significantly lower MAE. This result makes logical sense as it's more likely we are similar in opinion to a person who we share interests with than a person who we share friends with. However, the problem with item-rating information is the information that we lack when cold-starting a user.  

In the challenging case of cold start users, the algorithms that only use the Trust and Intra-Item information would be very beneficial, but these algorithms, unsuperisingly, perform not as well. Among these, we found that the Jaccard WA approach provided the best performance, outperforming random walk and intra-item-based approaches. A further idea that could be explored is simply using an approach like the Universal Random, which was one of our baseline recommenders. This recommender performed reasonably when averaged across our data sets and predicts preferences based on the distribution of current ratings for said item. 
 
While combining information, we found it wasn't easy to outperform the Item-Jaccard-based recommender.
The one case where we managed to improve upon the base Item-Jaccard model was when we performed addition with the Jaccard Index. So the similarity between the two users is based on the sum of the Jaccard Index and Item-Jaccard Index. We think that the performance of the recommender is increased as the Jaccard provides some marginal information which can impact the ratings in the case when a user trusts a celebrity. That way the user has a non-zero Jaccard index with many other nodes who also trust said celebrity, hence, slightly shifting the rating prediction in the direction of these users. 

We attempted combination models of various forms, combining the similarity scores with multiplication, summation, and a maximum function. However, we discovered that scaling the distributions had minimal impact on the resulting performance. We also concluded that simply using an equally weighted summation of the two scores leads to the simplest and most effective model. Hence, the highest-performing recommendation model combined trust information and item rating information. Though, it is not suitable for cold-start users.

We also found that recommenders based on the WA framework outperformed random-walk methods. Furthermore, the WA framework outperforms a similar approach called Majority of Majorities. We believe this approach to collaborative filtering is effective because it is not constrained by the structure of the graph (as is a random walk) and can produce fine-grained recommendations that can be non-integer values. Whereas, the MoM approach forces a specific integer rating to be assigned and hence is not as representative of the user's preference. 

When it came to the intra-item information, we found it performed quite poorly from the perspective of $\mu_{MAE}$. However, an interesting observation, as per Table~\ref{tab:merged_and_sorted_results}, was that the algorithms that included intra-item information were the most consistent in their performance—featuring the lowest $\sigma_{MAE}$ across all data sets. Thus, it can be deduced that the intra-item information is additive from a stability perspective, making a recommender perform with similar accuracy for all users. This idea was reinforced when we created the fully combined model, which utilized all information types. This was the ``Jaccard Item-Jaccard JII Combination WA'' model. We found this model performed worse than the Jaccard Item-Jaccard WA model in $\mu_{MAE}$, however, it performed better in mean $\sigma_{MAE}$. Hence, in a scenario where providing consistently good predictions for all users is of importance, the introduction of intra-item information could facilitate this.

Through our experimentation with opposing intra-item similarity metrics, we determined that the Intra-Item Jaccard approach outperformed the Pearson Correlation metric on downstream tasks. This was determined by building models upon these metrics and comparing the resulting accuracy. When further testing with the Intra-Item Jaccard similarity metric was undertaken, the resulting models performed better than random, and in combination led to stable models as described earlier. A recurring theme in these experiments was the very impressive performance of the Jaccard Index when used in a variety of applications. It was the go-to for drawing similarity scores using all information forms and is a simple but logical way to reason about the similarity of sets.

Testing the performance of our algorithms on an adversarial network provided some interesting insights. The most obvious of which is that item rating information is highly susceptible to an attack with fake accounts. The trust graph is relatively robust to adversaries, as one must influence individuals to shift the dynamics of this information. This is represented in the data by the recommendation systems involving the item rating information suffering from the most severe reductions in their MAE. We found that the least impacted algorithm was the MC Random Walk. As the algorithm propagates from a user, through the network to derive the rating, it will arrive at a legitimate node with a rating before venturing into the adversaries. The only case where the adversaries have an impact is when no ratings are found and the randomness means a rating is taken uniformly from the network. In these cases, the volume of ratings from adversaries impacts results. The next best algorithm for being robust in the face of adversaries was Jaccard WA. This is a Weighted Average based entirely on the Jaccard Index. This algorithm performs better due to the value it places on the trust graph, very rarely do nodes have a non-zero Jaccard similarity with the bad actors, hence the high performance.

It is very important to note that the goal of our study is not focused on surpassing state-of-the-art techniques; our study is intended to be a methodical investigation of the function of different factors in recommendation systems. Numerous sophisticated methods, like deep learning-based strategies, make use of complex architectures that maximize efficiency but frequently mask the distinct contributions of various elements that can be extracted from the social networks. We allow for a controlled study of the impact of incremental information on recommendation quality by keeping the setting simple. This allows us to isolate and better understand the contribution of each parameter, which would be challenging in highly complex setups. Interestingly, our model also performs comparatively well against other methods, showing the power of graph structure information that are often ignored, while being very simple and scalable.

The computational complexity of various similarity metrics and recommendation methods is analyzed in terms of the number of users (\(N\)), the number of neighbors for a given user (\(n\)), the number of items reviewed by a user (\(i\)), and the complexity of a similarity metric (\(s\)). The complexity of computing a metric for one pair of items or users varies: the Jaccard Index has a complexity of \(O(n)\), the Item-Jaccard Index and Item Rating Difference both have a complexity of \(O(i)\), while the Intra-Item Jaccard requires \(O(N)\) computations. When computing recommendations for a single user, the Weighted Average approach operates at \(O(N*s)\), the Mean/Median of Neighbors method has a complexity of \(O(n)\), and the Random Walk Monte-Carlo method scales as \(O(\frac{1}{\alpha})\). These methods are much faster than sophisticated collaborative filtering methods like LightGCN \cite{he2020lightgcn} or Neural Collaborative Filtering \cite{he_liao_zhang_nie_hu_chua_2017}.

\textbf{Future Work:} In investigating intra-item information, we found that the "Intra-Item Jaccard" metric was more effective than Pearson Correlation in determining item similarity. This result suggests further exploration of the Trust-Walker~\cite{10.1145/1557019.1557067}, which may lead to improved performance. 

Further research is needed to improve the computational efficiency of the WA approach. Unlike, random-walk methods, the WA-based recommenders require iterating over all nodes in the graph. Optimizations such as vectorization or caching similarities between users could be explored.

Furthermore, this work focused on fake accounts and fake ratings, but other attacks such as bribing popular nodes or adversarial censorship should be investigated in future studies. 

\section*{Authors' Contribution}
\textbf{Paras Stefanopoulos}: Conceptualization, Formal Analysis, Methodology, Software, Validation, Writing - Original draft preparation. \textbf{Sourin Chatterjee}: Software, Validation, Writing - Original draft preparation. \textbf{Ahad N. Zehmakan}: Conceptualization, Methodology, Supervision, Validation, Writing - Review. 

\section*{Declaration of Competing Interest}
The authors declare that they have no known competing financial interests or personal relationships that could have appeared to influence the work reported in this paper.

\end{document}